\documentclass[aps,showpacs,twocolumn,floatfix,amsmath,preprintnumbers,nofootinbib,a4paper,secnumarabic]{revtex4} 

\usepackage{amsmath,amssymb,hyperref,url}
\usepackage{epsfig,verbatim,amsfonts,graphicx}

\usepackage[caption=false]{subfig}
\def\sc{0.33}
\def\sc{0.28}
\def\sct{0.924} 

\def\art{paper}

\def\jrn#1#2#3#4#5#6{\textit{#3} \textbf{#4}, #5 (#6)} \def\boo#1#2#3#4#5#6#7{\textit{#2} (#3, #4, #5)}    \def\andd{ and } \def\andt{ and } \def\eq{eq.\,}  \def\Fig{figure~} \def\Figs{figures~}    \def\Sec{Sec.} 

\def\scn#1#2{\section{#1}\lb{#2}} 

\def\bfl{\begin{flushleft}}
\def\efl{\end{flushleft}}
\def\bfr{\begin{flushright}}
\def\efr{\end{flushright}}
\def\bc{\begin{center}}
\def\ec{\end{center}}
\def\be{\begin{equation}}
\def\ee{\end{equation}}
\def\bse{\begin{subequations}}
\def\ese{\end{subequations}}
\def\ba{\begin{eqnarray}}
\def\ea{\end{eqnarray}}
\def\baa#1{\begin{array}{#1}}
\def\eaa{\end{array}}
\def\bw{\begin{widetext}}
\def\ew{\end{widetext}}
\def\nn{\nonumber }
\def\lb#1{\label{#1}}
\def\bit{\begin{itemize}}
\def\eit{\end{itemize}}
\def\bco{}
\def\bcs{\begin{cases}}
\def\ecs{\end{cases}}

\def\Der#1#2{\frac{\drm #1}{\drm #2}}

\def\Sin#1#2{\, \text{sin}^{#1}#2}

\def\cf{\eta}

\def\vol{{\cal V}}

\def\Mass{{\cal M}}

\def\vena{\boldsymbol{\nabla}}
\def\ve#1{\boldsymbol{#1}}

\def\drm{d}
\def\dvol{\drm\vol}

\def\dn{\rho}  \def\dnc{\bar{\dn}}
\def\lnc{\bar{\ell}}

\def\tfc{q}
\def\lapl{\vena^2}

\def\nc0{b_0}

\def\wfv{\Psi_{\text{vac}}}

\def\grpot{\Phi}
\def\grpott{\grpot} 
\def\grsmi{\grpot_{\text{smi}}}
\def\grmic{\grpot_{\text{RN}}}
\def\grnew{\grpot_{\text{N}}}
\def\grgal{\grpot_{\text{(ln)}}} 
\def\grmgal{\grpot_{\text{(1)}}} 
\def\grcos{\grpot_{\text{(2)}}} 

\def\vgal{v_{\text{(ln)}}}

\def\rad{r}
\def\rve{\ve{r}}
\def\pwr{\chi}

\def\ror{R}

\def\mesph{\sigma^2} 
\def\sun{\odot}
\def\m#1{\multicolumn{1}{c}{$ #1$}}

\begin{document}

\preprint{\small Pramana \textbf{97}, 2 (2023) 
\quad [\href{https://doi.org/10.1007/s12043-022-02480-2}{DOI: 10.1007/s12043-022-02480-2}]}

\title{Galaxy rotation curves in superfluid vacuum theory}

\author{Konstantin G. Zloshchastiev}
\email{http://orcid.org/0000-0002-9960-2874} 
\affiliation{Institute of Systems Science, Durban University of Technology,
P.O. Box 1334, Durban 4000, South Africa\\
E-mail: kostiantynz@dut.ac.za, kostya@u.nus.edu}

\begin{abstract}
Logarithmic superfluid theory of physical vacuum suggests that gravity has a multiple-scale structure;
where one can recognize sub-Newtonian, Newtonian, logarithmic, linear and quadratic (de Sitter) terms in the induced spacetime
metric and effective potential. To test the theory's predictions on a galactic scale,
we apply best-fitting procedures to the rotation curve data obtained from fifteen galaxies by the HI Nearby
Galaxy Survey; assuming their stellar disk's parameters to be fixed to the mean values measured 
using photometric methods. Although the fitting results seem to be sensitive to a stellar disk model chosen, they correspond closely with observational data,
even for those galaxies which rotation velocity profiles do not have flat regions.
\end{abstract}


\date{received: 24 Feb 2022} 



\maketitle

\scn{Introduction}{s:intro}
According to Newtonian gravity, the rotation velocity of non-relativistic bodies is inversely proportional to the square root of the distance from a gravitating center, thus resulting in Keplerian orbits.
This theory is expected to be applicable to rotating stars and gas in galaxies, especially far from galactic cores where relativistic effects become negligible.
However, multiple observations of rotation velocities of stars around 
the centers of galaxies reveal tremendous discrepancies from Keplerian behaviour.
These grow as one proceeds from the inner galactic regions to the furthest data points. 
The observed data shows rather diverse behaviour of rotation curves, often including, but not restricted to, flat rotation curve (FRC) regimes in the outer regions. 
Generally, there are two most popular ways of addressing this problem. 

The most direct way is that there is a large amount of non-luminous matter, dubbed dark matter (DM), which interacts with luminous matter only through gravity. 
However, after decades of research, no direct evidence of the existence of dark matter particles has been found. 
Besides, Standard Model of particle physics and related experiments have placed rather strict bounds upon exotic particles' existence and properties.
This inspires further theoretical studies in search of other explanations for the discrepancy between 
luminous matter's gravitating potential and its observed one.

There is also a growing understanding that a convincing theory of DM-attributed phenomena cannot 
be a stand-alone or purely phenomenological model; but should, instead, be rooted in an axiomatic fundamental theory involving all known interactions.
It is also agreed  that creating this fundamental theory would not be possible without
a clear picture of what the physical vacuum - a natural phenomenon 
which underlies all known particles and interactions - is all about.

One of the simplest and physically clear candidates for a theory of physical vacuum and quantum gravity 
is superfluid vacuum theory (SVT).
This framework has foundations in Dirac's idea of 
the physical vacuum being a nontrivial object described by quantum wavefunction \cite{dir51}.
Since his work,
various approaches have been proposed,
which agree on the main paradigm
(of physical vacuum being a background quantum liquid
and elementary particles being excitations thereof \cite{volbook,huabook}),
but which vary in details. 
In this \art, we will be using a modern version of the superfluid vacuum approach,  
logarithmic superfluid vacuum theory;
its details can be found in the landmark papers \cite{z10gc,z11appb,z20un1}.

The \art~ is organized as follows.
Theory of physical vacuum based on the logarithmic superfluid model
is outlined in \Sec~\ref{s:mod}, 
including the derivation of effective gravitational potential
induced by the superfluid vacuum in a given quantum state.
Galaxy rotation curves and their best fit results based on the logarithmic superfluid vacuum theory are presented 
in \Sec~\ref{s:frc}.
Discussion of results and conclusions are presented in sections \ref{s:dis} and \ref{s:con}, respectively.

\scn{Induced spacetime and effective gravity 
}{s:mod}
In this section, closely following the lines of the work \cite{z20un1},
we derive the effective gravitational potential
of the logarithmic superfluid vacuum in the presence of a large density inhomogeneity,
serving as a gravitating center and reference point.

The theory assumes that 
the physical vacuum is described,
when disregarding
quantum fluctuations, by 
the fluid wavefunction $\Psi (\rve,\, t)$,
which is a three-dimensional Euclidean scalar.
The quantum state of the fluid itself is described by a ray in the corresponding Hilbert space,
therefore,
this wavefunction obeys the normalization condition
$ 
\langle \Psi | \Psi \rangle
= \int_\vol \dn \, \dvol 
= \Mass
$, 
where 
$\Mass$ and $\vol$  
are the
total mass and volume of the fluid, respectively,
and
$\dn = |\Psi|^2$ is the fluid mass density.
The wavefunction's evolution is governed by a wave equation of  
a Schr\"odinger type but with logarithmic nonlinearity:
\be
i\hbar \partial_t \Psi
=
\left[
-\frac{\hbar^{2}}{2 m} \lapl
+V_\text{ext}  (\rve, t)
-
\left(
b_0 - \frac{\tfc}{\rad^2}
\right)\!
 \ln{\!\left(\frac{|\Psi|^{2}}{\dnc} \right)}
\right]\!\Psi
,\label{e:vcm}
\ee
where 
$m$ 
is the constituent particles' 
mass,
$V_\text{ext} (\rve,t)$ is an external or trapping potential (to be neglected for simplicity in what follows),
$\rad  = |\rve| = \sqrt{\rve \cdot \rve}$
is a radius-vector's absolute value,
and $b_0$ and $q$ are real-valued constants.
Despite its resemblance to Schr\"odinger equations,
\eq\eqref{e:vcm} should not be confused with theories of nonlinear quantum mechanics \cite{bb76}.
Instead, the logarithmic nonlinearity is induced by many-body effects in the quantum Bose liquid,
whereas the underlying quantum mechanics remains conventional.  

In this picture, 
massless excitations, such as photons, are somewhat analogous to  
acoustic waves propagating with velocity 
$ 
 c_s \propto \sqrt{p' (\dn)}
$, 
where 
fluid pressure is determined via the equation of state $p =p (\dn)$,
and the prime denotes a derivative with respect to the argument in brackets.
A relativistic observer
``sees'' himself located inside four-dimensional curved spacetime with a pseudo-Riemannian metric:
\be\lb{e:metr1}
g_{\mu\nu} 
\propto
\frac{\dn}{c_s}
\left(
\baa{ccc}
-\left[ c_s^2 - \cf^2 (\vena S)^2 \right] & \vdots & -  \cf \vena S \\
\cdots\cdots & \cdot & \cdots \\
-  \cf \vena S & \vdots & \textbf{I}
\eaa
\right)
,
\ee
where $\cf = \hbar/m$,
$S = S (\rve,\, t) = - i \ln{\left(\Psi (\rve,\, t) /|\Psi (\rve,\, t)|\right)}$ is a phase of the condensate wavefunction written
in the Madelung representation,
$\Psi = \sqrt\dn \, \exp{(i S)}$,
and
$\textbf{I}$ is a three-dimensional unit matrix.
In this picture, Einstein field equations 
are interpreted as a definition for an induced stress-energy tensor,
describing
some effective matter, as observed by a relativistic observer,
$ 
\widetilde{T}_{\mu\nu} 
\equiv
\kappa^{-1}
\left[
R_{\mu\nu} (g) - \frac{1}{2} g_{\mu\nu} R (g)
\right]
$, 
where $\kappa = {8 \pi G}/{c_{(0)}^2}$ is the Einstein's gravitational constant,
$G$ is the Newton's gravitational constant, 
and
$ c_{(0)} \approx c $,
where  
$c = 2.9979 \times 10^{10}$ $\text{cm}\, \text{s}^{-1}$
is historically
called the \textit{speed of light in vacuum}.


If the physical vacuum is in a 
quantum state described
by wavefunction $\Psi = \wfv (\rve,\, t)$,
then the solution of \eq \eqref{e:vcm} is equivalent to the solution
of the linear Schr\"odinger equation
with effective potential,
which can be written (in Cartesian coordinates) as:
\be\lb{e:igrav}
\grpott 
=- \frac{1}{m} V_\text{eff} (\rve, t) 
=
\frac{1}{m} 
\left(
b_0 - \frac{\tfc}{\rad^2}
\right)
\ln{\!\left(\frac{|\wfv (\rve, t)|^{2}}{\dnc} \right)}
,
\ee
where we disregard any anisotropy
and rotation.

According to logarithmic SVT,
Lorentz symmetry emerges
in the ``phononic'' (low-momentum) limit, 
where SVT can be well approximated  by the theory of general relativity \cite{z20ijmpa,z21ltp}.
Correspondingly, a relativistic observer 
would perceive gravity induced by the potential \eqref{e:igrav}
as curved four-dimensional spacetime.
In a rotationally invariant case,
the line element of such spacetime
can be written as
\ba
d s^2 
&=&
- K^2 c_{(0)}^2  d t^2
+ \frac{1}{K^2} d r^2 + 
R^2 (r) d \mesph
,\lb{e:metrst}\\
K^2 &\equiv& 1 + 2 \grpott  /c_{(0)}^2
,\nn
\ea
where 
$R (r) = r \left[1 + {\cal O} \left(\grpott /c_{(0)}^2\right)\right] \approx r$,
$d\, \mesph = d\theta^2 + \Sin{2}{\theta} \, d \varphi^2$
is the line element of a unit two-sphere.

Furthermore, because we do not know the exact wavefunction of the physical vacuum,
we must resort to general arguments and trial functions whose parameters 
are to be determined empirically. 
We expect that our vacuum is currently in a stable state,
which is close to a ground state 
or at least to a metastable state, with a sufficiently large lifetime.  
It is thus natural to assume that the state 
$|\wfv \rangle$
is stationary and rotationally invariant,
and use the following ansatz for its amplitude:
\be\lb{e:avex}
|\wfv | =
\sqrt{\dnc}\,
\left(
\frac{\rad}{\lnc}
\right)^{\pwr/2}
P 
\left(
\rad 
\right)
\,
\exp{\!
\left(
-
\frac{a_2 }{2} \rad^2
+
\frac{a_1}{2} \rad
+ 
\frac{a_0}{2}
\right)
}
,
\ee
where 
$P (\rad)$ is a  polynomial function,
$\pwr$
and $a$'s are constant parameters of the solution.
Here $\lnc$ is the length scale of the logarithmic nonlinearity -- 
which can be chosen to be
the classical  length   $(m/\dnc)^{1/3}$
or the quantum temperature length
$\hbar/\sqrt{m b_0}$.
For reasons explained in
\cite{z20un1}, 
these constants' values
remain theoretically unknown at this stage.
From the empirical point of view, the function \eqref{e:avex} can be considered  a
trial function, whose parameters can be fixed using experimental data.

By
substituting the trial solution \eqref{e:avex} into the definition \eqref{e:igrav},
we derive the effective gravitational potential
\ba
\grpott 
&=&
\grsmi
(\rad)
+
\grmic
(\rad)
+
\grnew
(\rad)
\nn\\&&
+
\grgal
(\rad)
+
\grmgal
(\rad)
+
\grcos
(\rad)
+
\grpott_0 
,\lb{e:grev}
\ea
where
$
\grpott_0 
=
\left(
a_0 b_0 
+ a_2 \tfc
\right)/m
$ 
is the additive constant,
\ba
\grsmi (\rad) 
&=&
-\frac{q}{m \rad^2}
\ln{\left[
\left(
\frac{\rad}{\lnc}
\right)^{\pwr}
P^2\! 
\left(
\rad
\right)
\right]}
, ~~~\lb{e:grsmi}\\
\grmic (\rad) 
&=&
-\frac{a_0 \, q}{m}
\frac{1}{\rad^2}
\, , \lb{e:grmic}\\
\grnew (\rad) 
&=&
-\frac{a_1 q}{m} \frac{1}{\rad} =
- \frac{G M}{\rad}
, \lb{e:grnew}\\
\grgal (\rad) 
&=&
\frac{b_0}{m}
\ln{\left[
\left(
\frac{\rad}{\lnc}
\right)^{\pwr}
P^2\! 
\left(
\rad 
\right)
\right]}
,  \lb{e:grgal}\\
\grmgal (\rad) 
&=&
\frac{a_1 b_0}{m}
\rad
,  \lb{e:grmgl}\\
\grcos (\rad) 
&=&
-\frac{a_2 b_0}{m}
\rad^2
,    \lb{e:grcos}
\ea
where
$
M 
$
is the (induced) gravitational mass of the configuration.

The main 
simplifying assumptions and approximations underlying
the derivation of the  potential
are summarized in the Appendix B of 
\cite{z20un1}.
Note also that in the work \cite{z20un1} we used a simple approximate form of the non-exponential part of the 
wavefunction amplitude,
$
\left(
\rad/\lnc
\right)^{\pwr/2}
P 
\left(
\rad/\lnc
\right) \approx 
\left(
\rad/\lnc
\right)^{\pwr/2}
$,
which resulted in a simplified form of logarithmic terms in the effective potential and induced spacetime metric.
That approximation was sufficient for the illustrative purpose of explaining the FRC regime,
and was also robust for asymptotic studies of rotation curves in  
\cite{z21rc};
but for the purposes of this \art~ it is prudent to do not specify $P(r)$
in formula \eqref{e:avex} until the next section. 

Last but not least, it is important to emphasize the meaning of the term `effective potential' here.
From the construction of \eq\eqref{e:igrav}, it is clear that no potential exists in reality -- 
instead,
many-body
quantum-mechanical effects in the supefluid act in the manner we perceive as  gravitational potential.
One can imagine a (distant) classical analogy of this process:
during the diffusion process of a substance, 
its
individual molecules move from a region of higher concentration to a region of lower concentration,
as if they were driven by some macroscopic potential.
In our case, the logarithmic term is directly related to the Everett-Hirschman information entropy of the logarithmic superfluid,
cf. Section 3 of 
\cite{z20un1},
therefore, it is this entropy's evolution which induces ``thermodynamic'' force and associated ``potential''.

\scn{Rotation curves}{s:frc}
From this section onwards, we will work in units 
$G = 1$.
In a spherically symmetric case,
the
rotation velocity of stars orbiting 
with non-relativistic velocities
on a plane driven by gravitational potential \eqref{e:grev}, 
or, alternatively, along geodesics in the induced spacetime \eqref{e:metrst},
can be computed using a simple formula
$ 
v^2 
= 
\frac{1}{2}
r \,
\Der{}{r} 
(
c_{(0)}^2
K^2
)
\bigl|_{r= \ror}
= 
\ror \,
\grpott' (\ror)
$, 
where $v$ is the orbital velocity
and $\ror$ is the orbit's radius.
We assume that this formula is approximately valid in
the cylindrically symmetric case, which is a common assumption for studies of this kind.

Besides,
we will neglect those terms in \eq\eqref{e:grev} which decay faster than the Newtonian term at spatial infinity.
We thus obtain
\ba
\grpott (\rad)
=
\grnew
(\rad)
+
\grgal
(\rad)
+
\grmgal
(\rad)
+
\grcos
(\rad) 
,\lb{e:greva}
\ea
where the additive constant is omitted as well.
Using this truncated potential,
we obtain
\bw
\ba
v (\ror) 
&=&  
\left\{
\ror \Der{}{\ror} 
\left[
\grnew (\ror) + \grgal (\ror) + \grmgal (\ror)
+
\grcos
(\ror)
\right]
\right\}^{1/2}
= 
\sqrt{
v_\text{N}^2 
+ \vgal^2 
+
\tilde a_1 \ror
-
\tilde a_2 \ror^2
}
, \lb{e:ov0}
\ea
where 
\ba
\vgal^2 
&=&
\frac{b_0}{m}
\ror \,
\Der{}{\ror} 
\ln{\!\left[
\left(
\frac{\ror}{\lnc}
\right)^{\pwr}
P^2\! 
\left(
\rad 
\right)
\right]}
= 
\frac{b_0}{m}
\ror \,
\Der{}{\ror} 
\left\{
\pwr
\ln{\!\left(
\frac{\ror}{\lnc}
\right)}
+
\ln{\!\left[
P^2\! 
\left(
\rad 
\right)
\right]}
\right\}
, 
\ea
where the tilde  denotes the multiplication by $b_0/m$
(such as $\tilde a_1 = (b_0/m) a_1$ {et cetera}),
and 
$v_\text{N}$ 
is the rotation velocity derived from Newtonian dynamics.

As mentioned in the previous section, an exact form of
the polynomial $P (x)$ is not yet theoretically known.
In the works \cite{z20un1,z21rc}, we used the approximation $P (r) \approx r$,
which was sufficient for our purposes there.
In this \art, 
we assume a slightly more general form
\be
P(r) \approx k_2 r^2 + k_1 r + 1
, 
\ee
therefore
\ba
\vgal^2 
\approx
\frac{b_0}{m}
\ror \,
\Der{}{\ror} 
\left\{
\pwr
\ln{\!\left(
\frac{\ror}{\lnc}
\right)}
+
\ln{\!\left[
\left(
k_2 \ror^2
+
k_1 \ror
+
1
\right)^2
\right]}
\right\}
, \lb{e:ovgal}
\ea
hence
\ba
v (\ror) 
\approx 
\sqrt{ 
v_\text{N}^2 
+ 
\frac{b_0}{m}
\ror \,
\Der{}{\ror} 
\left\{
\pwr
\ln{\!\left(
\frac{\ror}{\lnc}
\right)}
+
\ln{\!\left[
\left(
k_2 \ror^2
+
k_1 \ror
+
1
\right)^2
\right]}
\right\}
+
\tilde a_1 \ror
-
\tilde a_2 \ror^2
} 
, \lb{e:ov}
\ea
where the coefficients $k_i$ will also have their values fixed according to the best-fitting criteria.

Furthermore,
to calculate the velocity $v_\text{N}$,  the
mass distribution of a galaxy must be assumed,
usually derived from the photometric data. 
This generally contains 
two components: 
gas (mainly neutral hydrogen and helium,
any other gases are  negligible compared to them) and the stellar disk,
while the mass of the bulge is usually neglected for simplicity \cite{too63}. 
Also for a sake of simplicity, it is assumed that both gas and stellar disks are thin.

\begin{figure}[t]
\centering
\subfloat[DDO 154]{\includegraphics[width=\sc\columnwidth]{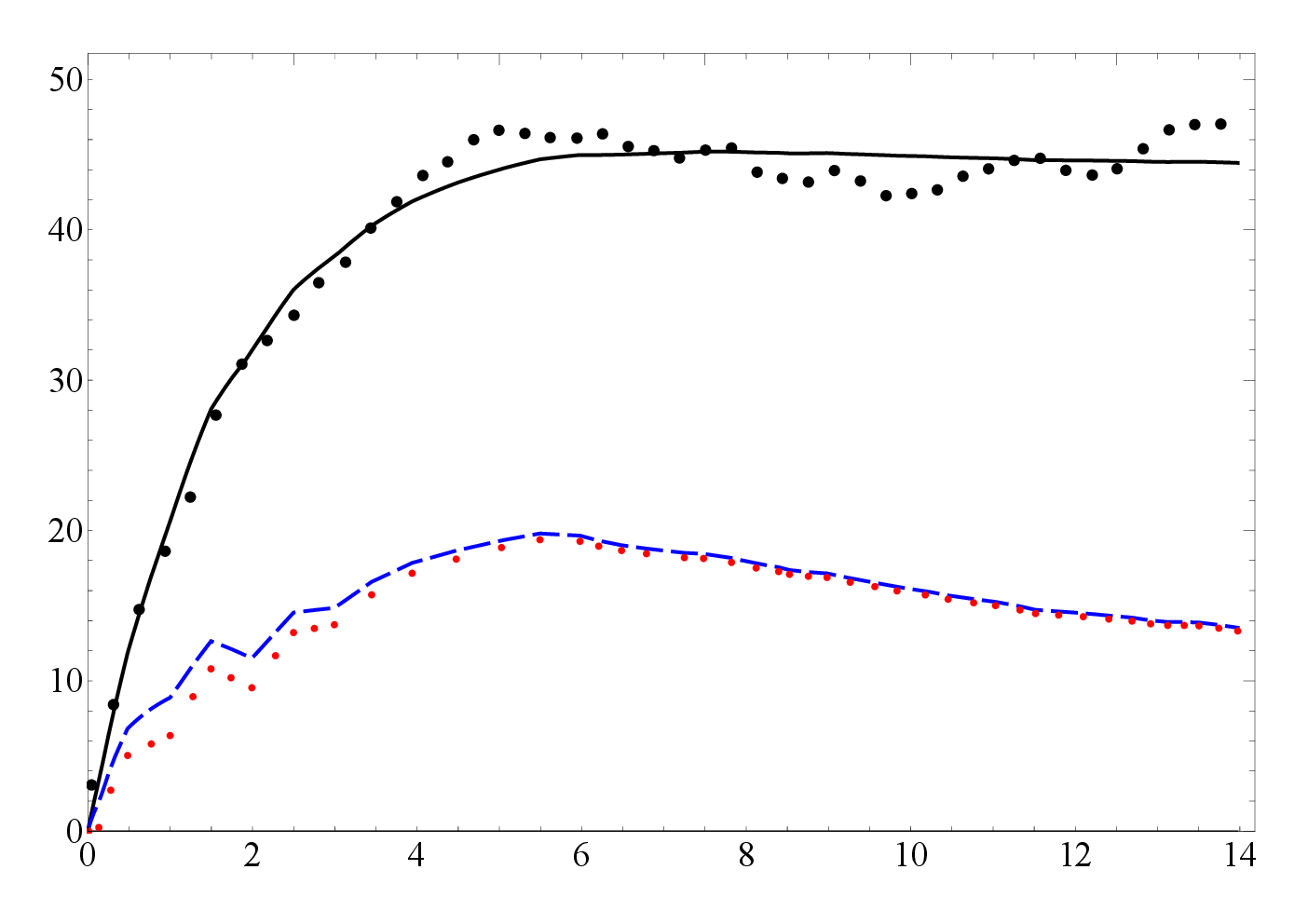}}
\subfloat[NGC 925]{\includegraphics[width=\sc \columnwidth]{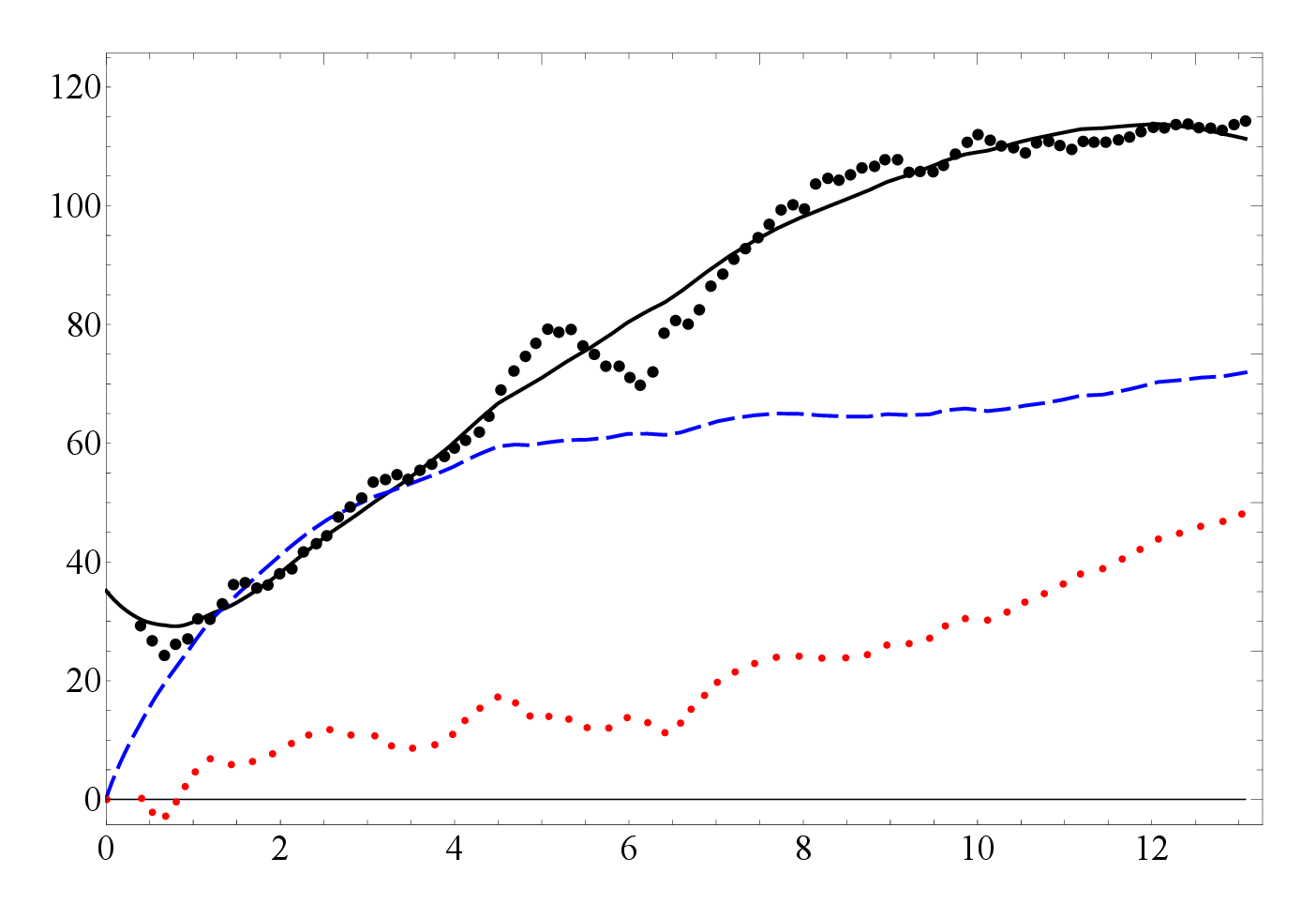}}
\subfloat[NGC 2403]{\includegraphics[width=\sc\columnwidth]{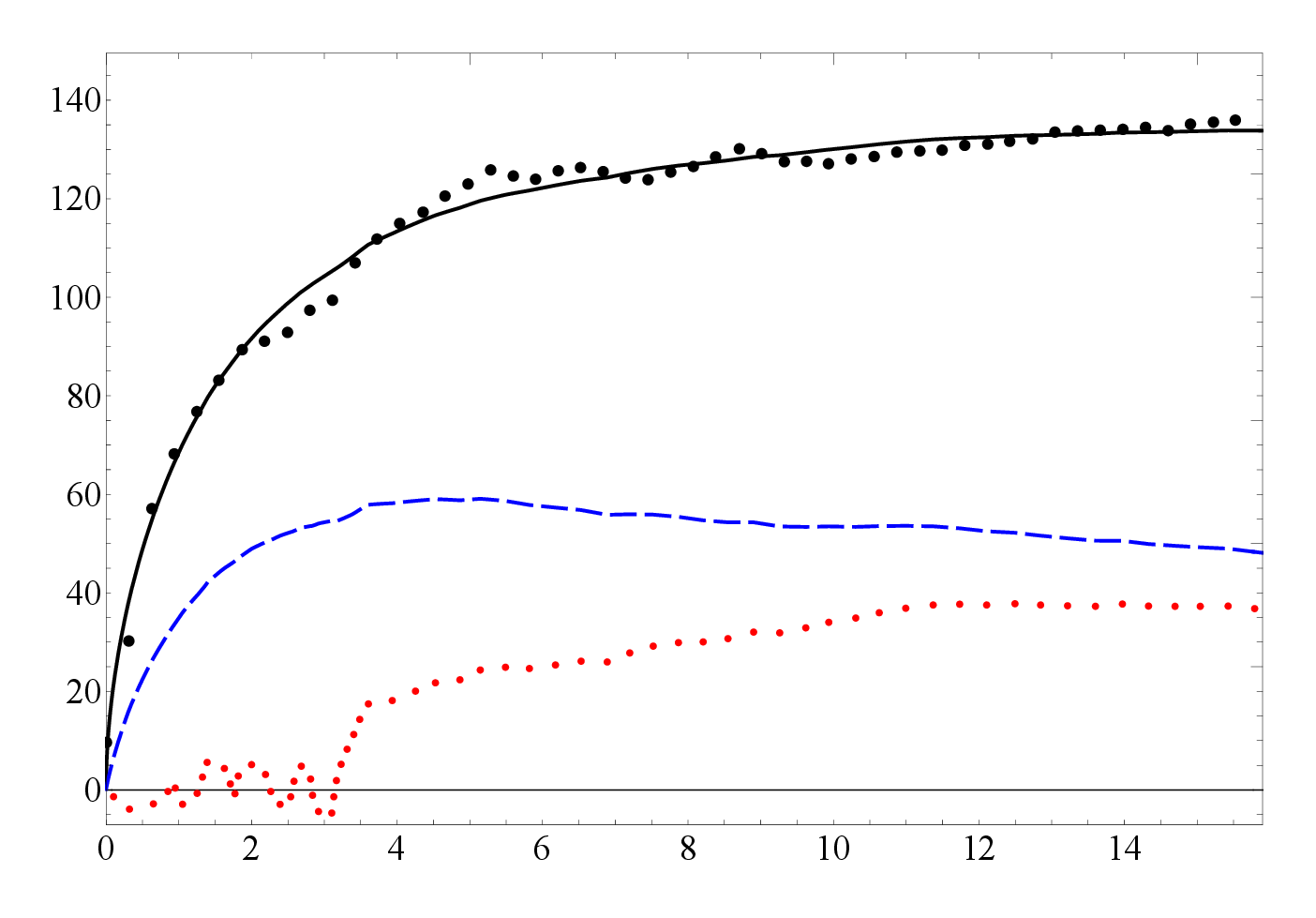}}\\
\subfloat[NGC 2841]{\includegraphics[width=\sc\columnwidth]{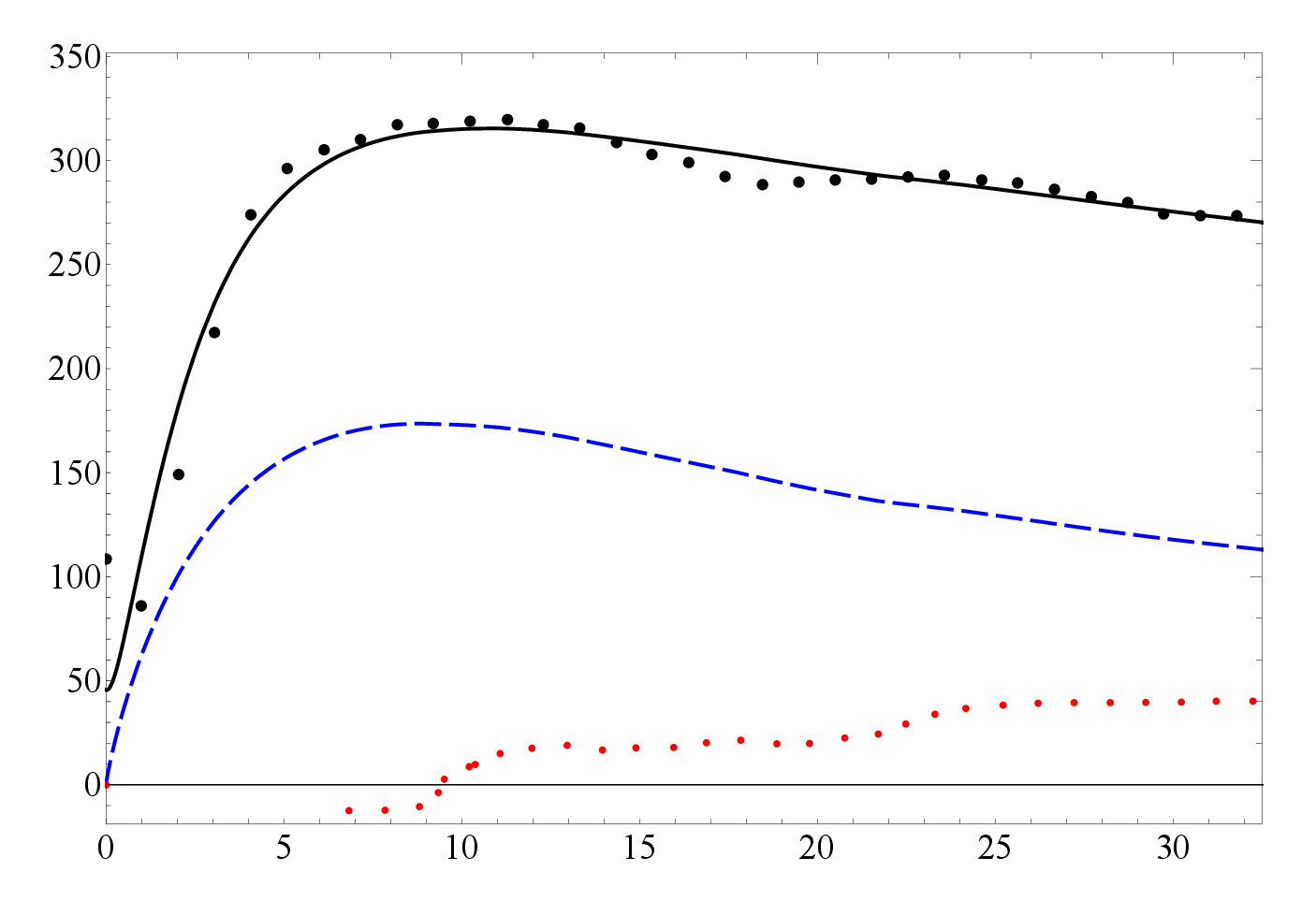}}
\subfloat[NGC 2903]{\includegraphics[width=\sc\columnwidth]{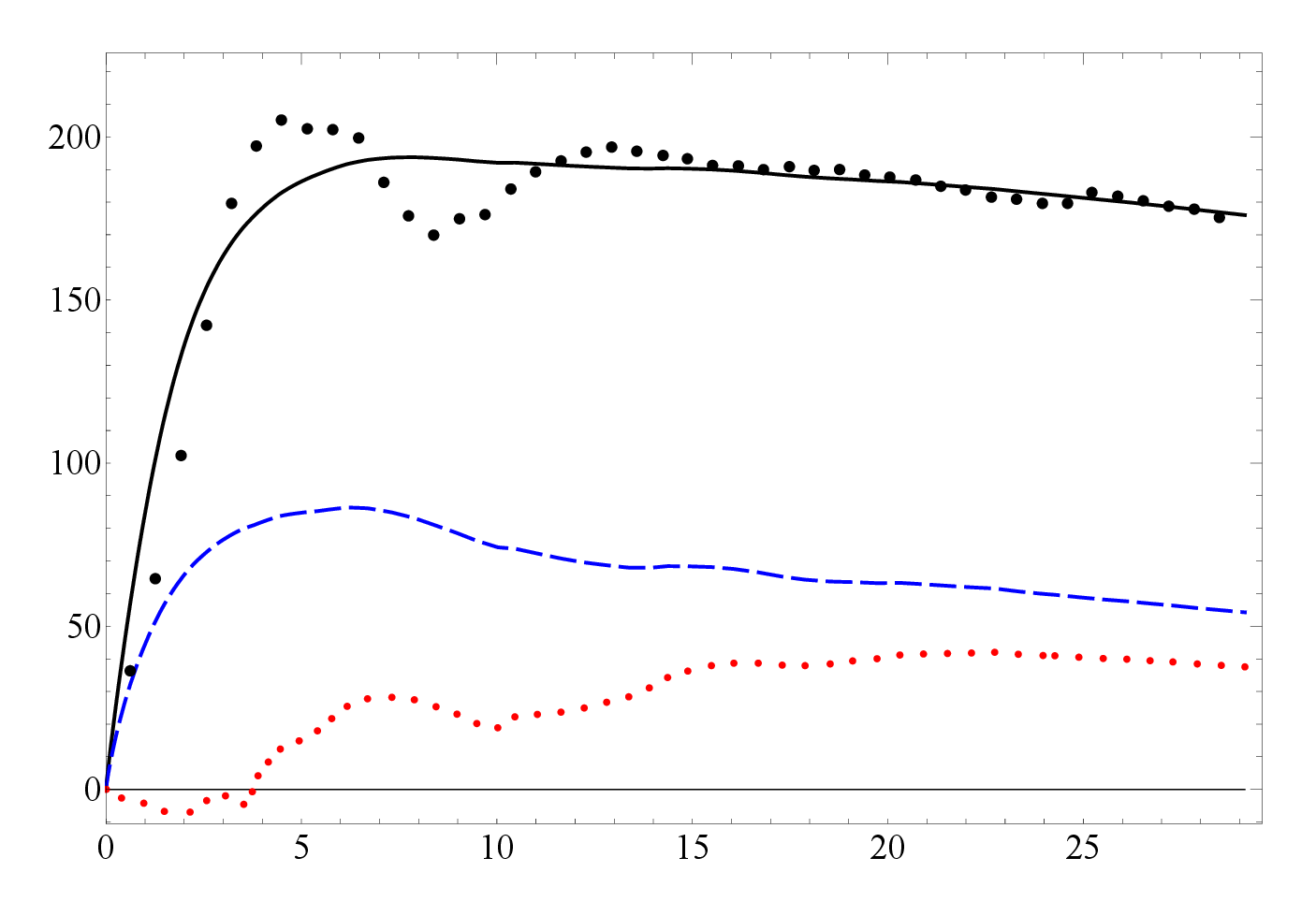}}
\subfloat[NGC 2976]{\includegraphics[width=\sc\columnwidth]{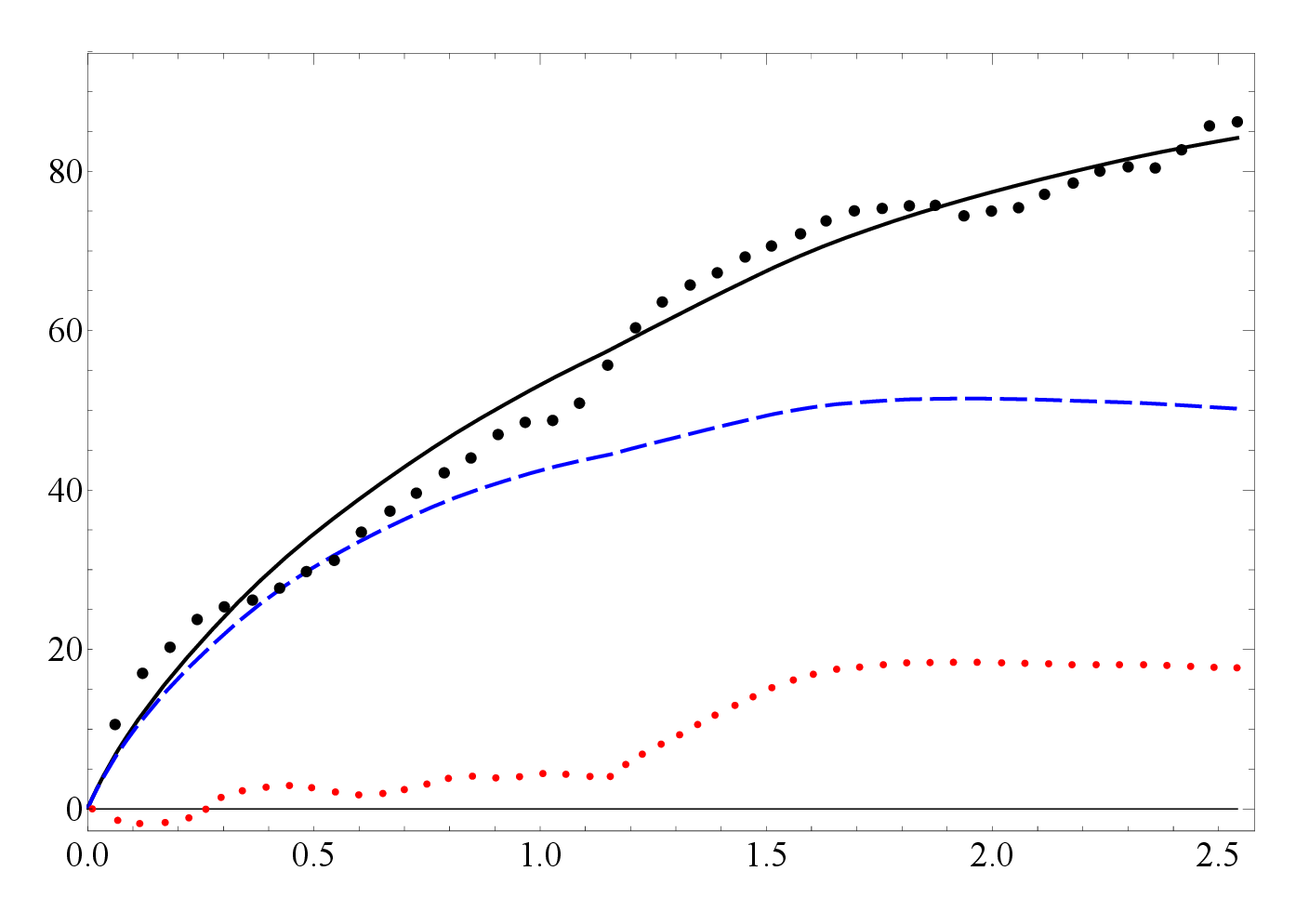}}\\
\subfloat[NGC 3031]{\includegraphics[width=\sc\columnwidth]{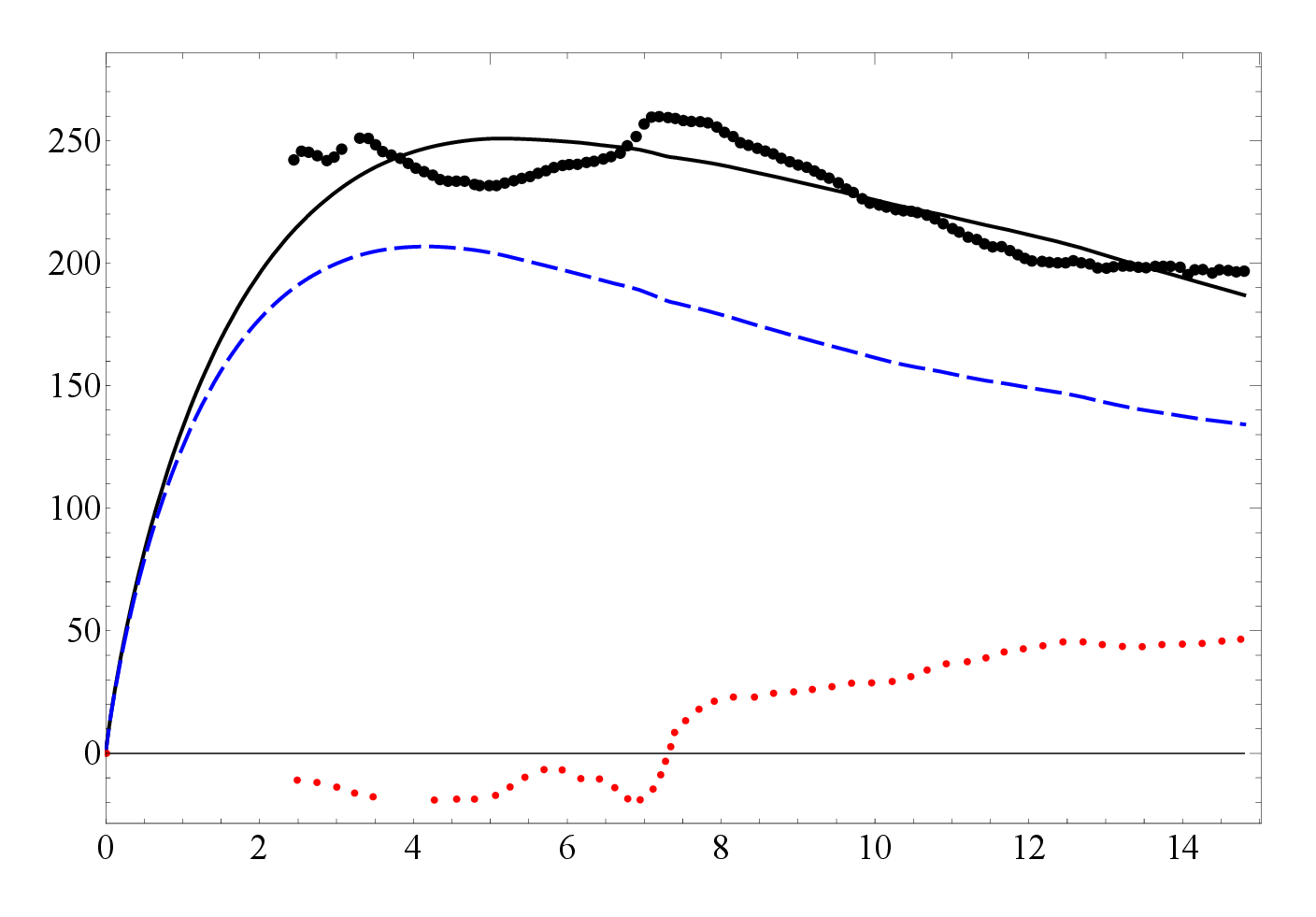}}
\subfloat[NGC 3198]{\includegraphics[width=\sc\columnwidth]{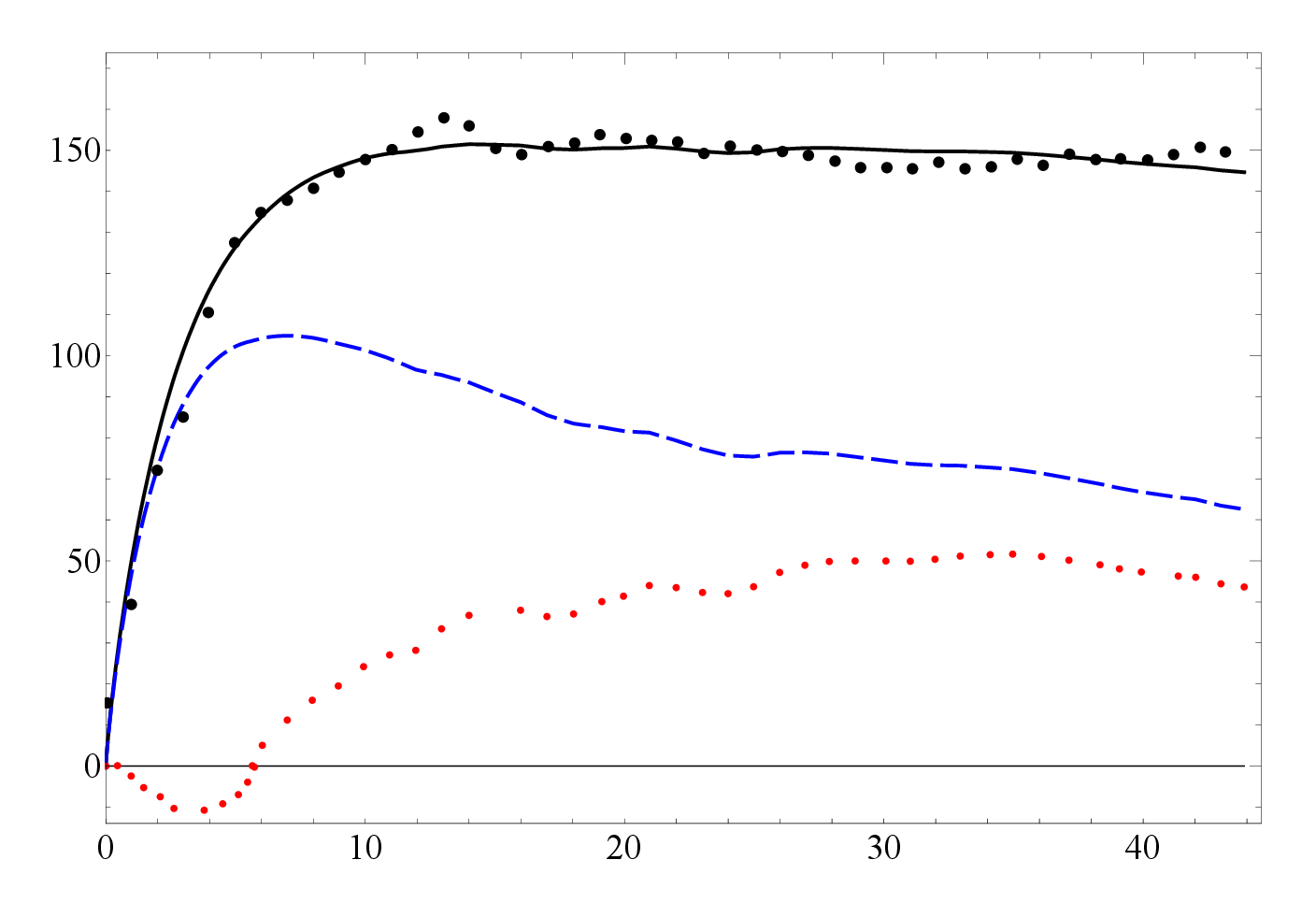}}
\subfloat[NGC 3521]{\includegraphics[width=\sc\columnwidth]{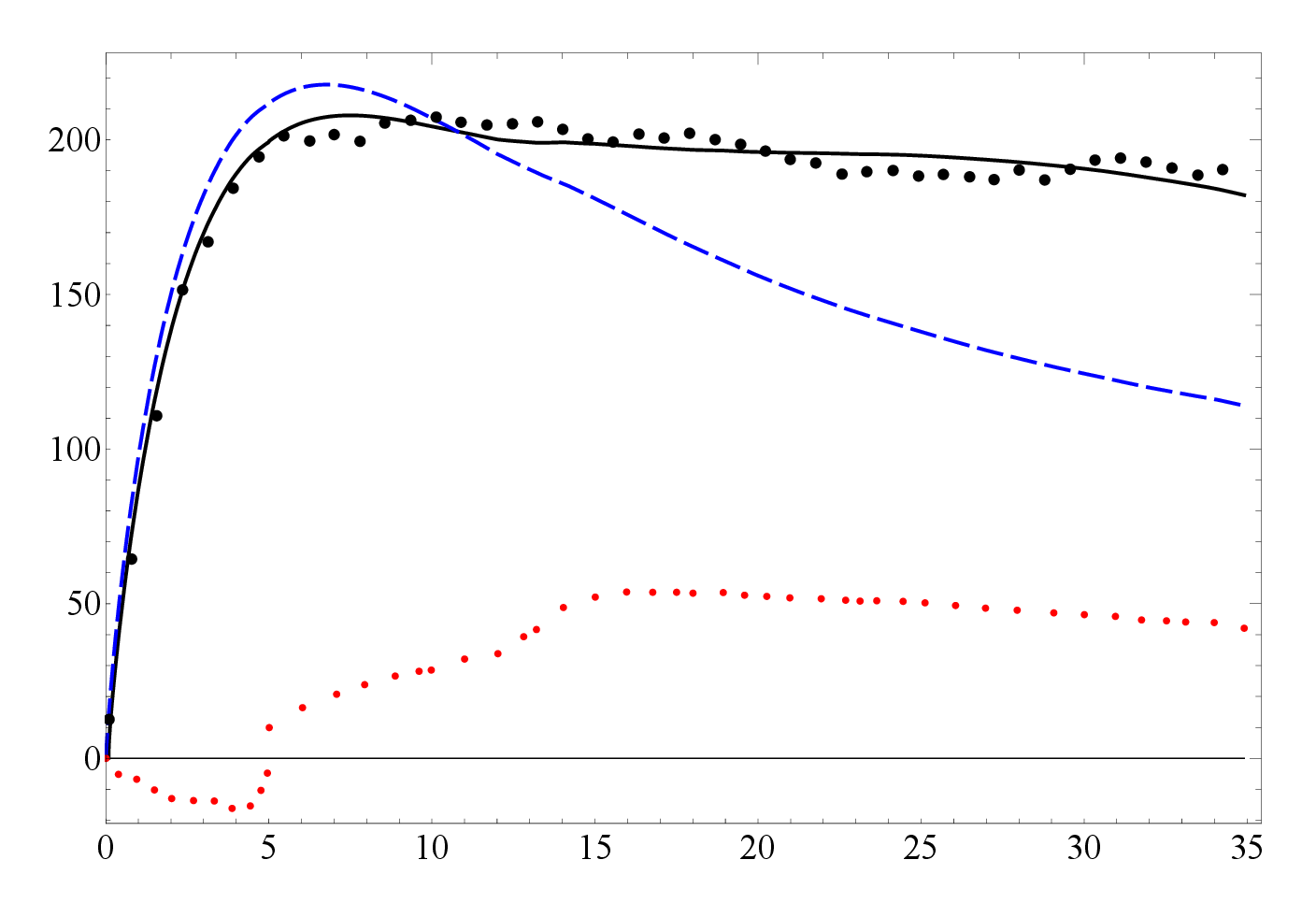}}\\
\subfloat[NGC 3621]{\includegraphics[width=\sc\columnwidth]{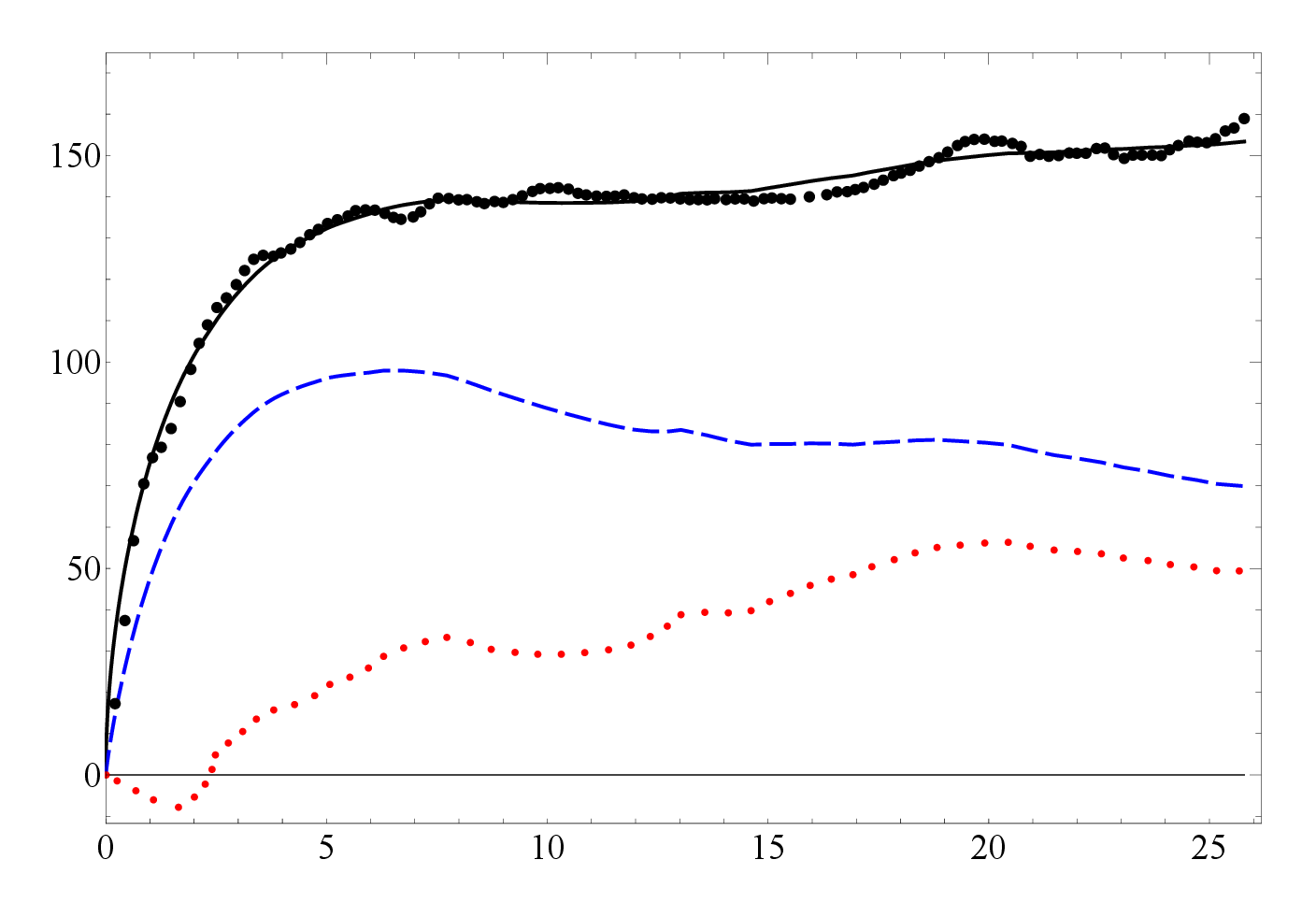}}
\subfloat[NGC 4736]{\includegraphics[width=\sc\columnwidth]{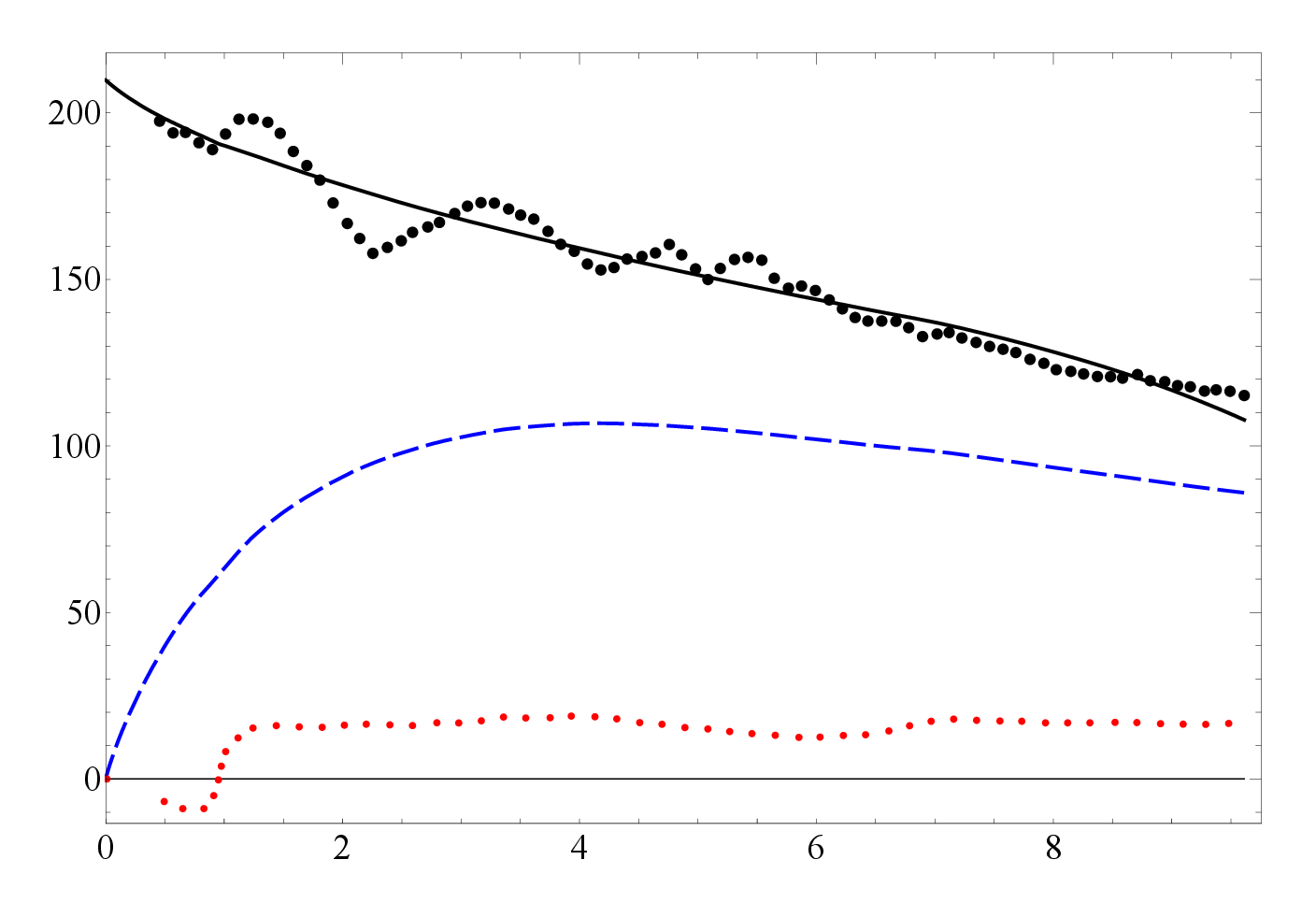}}
\subfloat[NGC 5055]{\includegraphics[width=\sc\columnwidth]{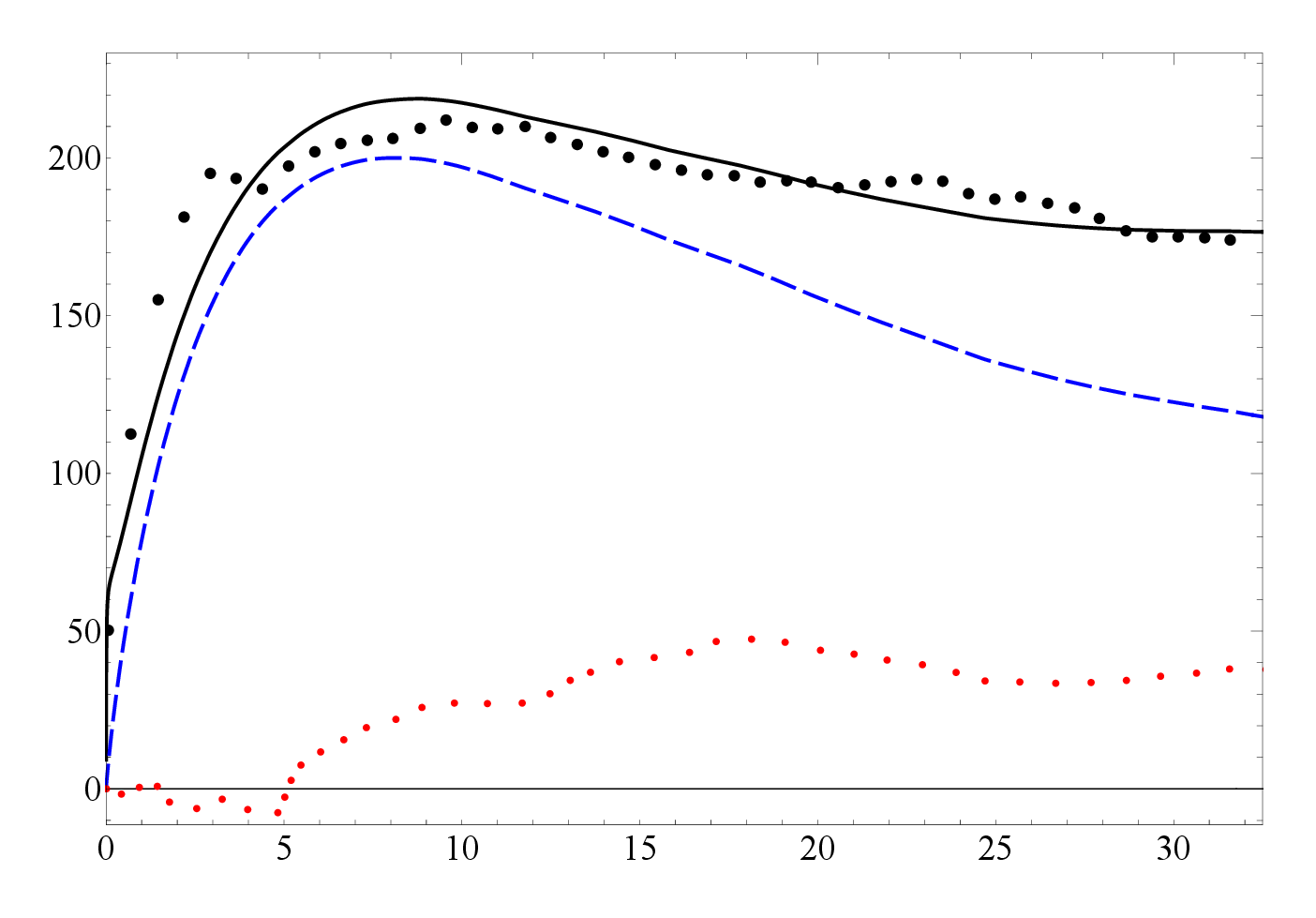}}\\
\subfloat[NGC 6946]{\includegraphics[width=\sc\columnwidth]{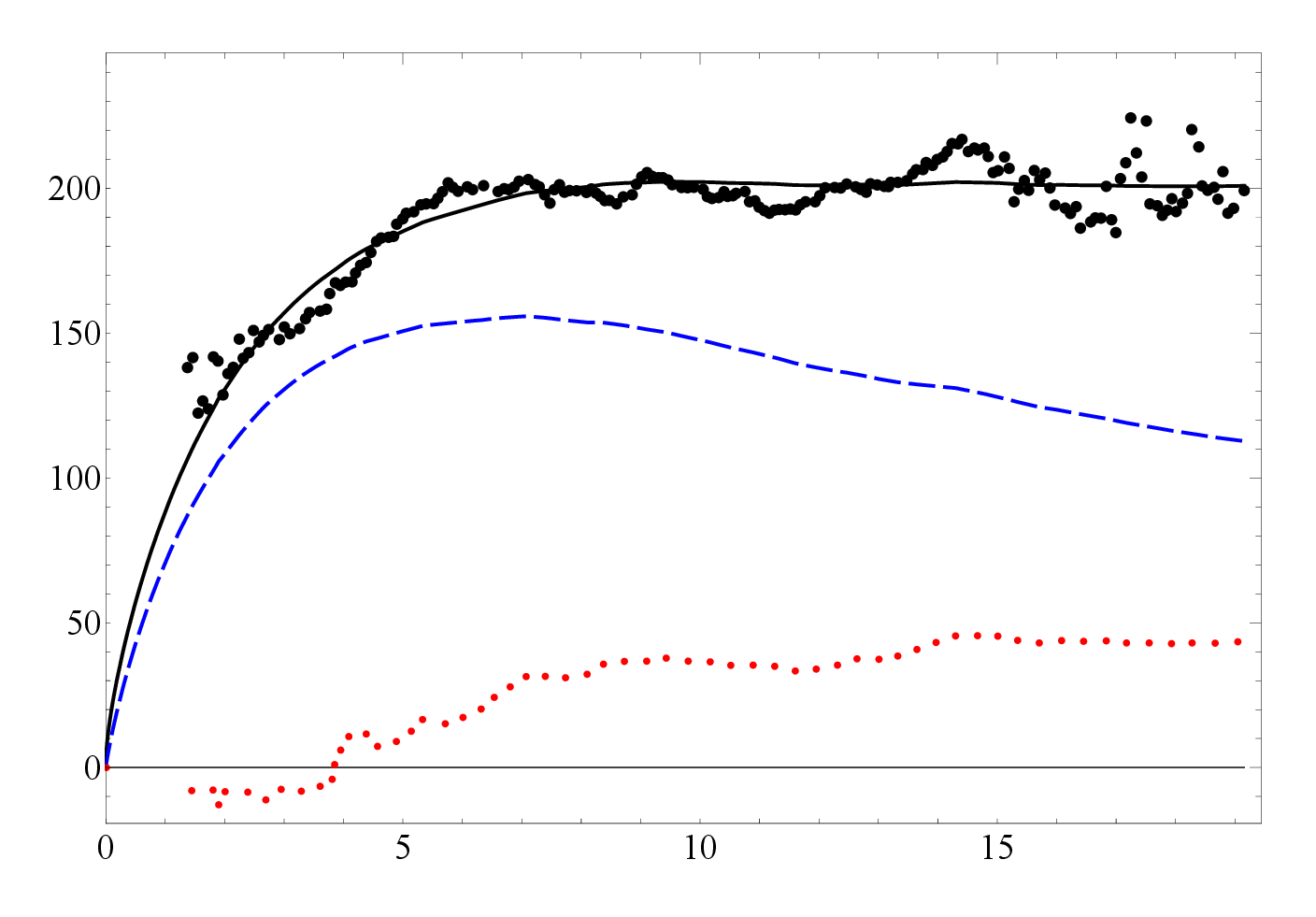}}
\subfloat[NGC 7331]{\includegraphics[width=\sc\columnwidth]{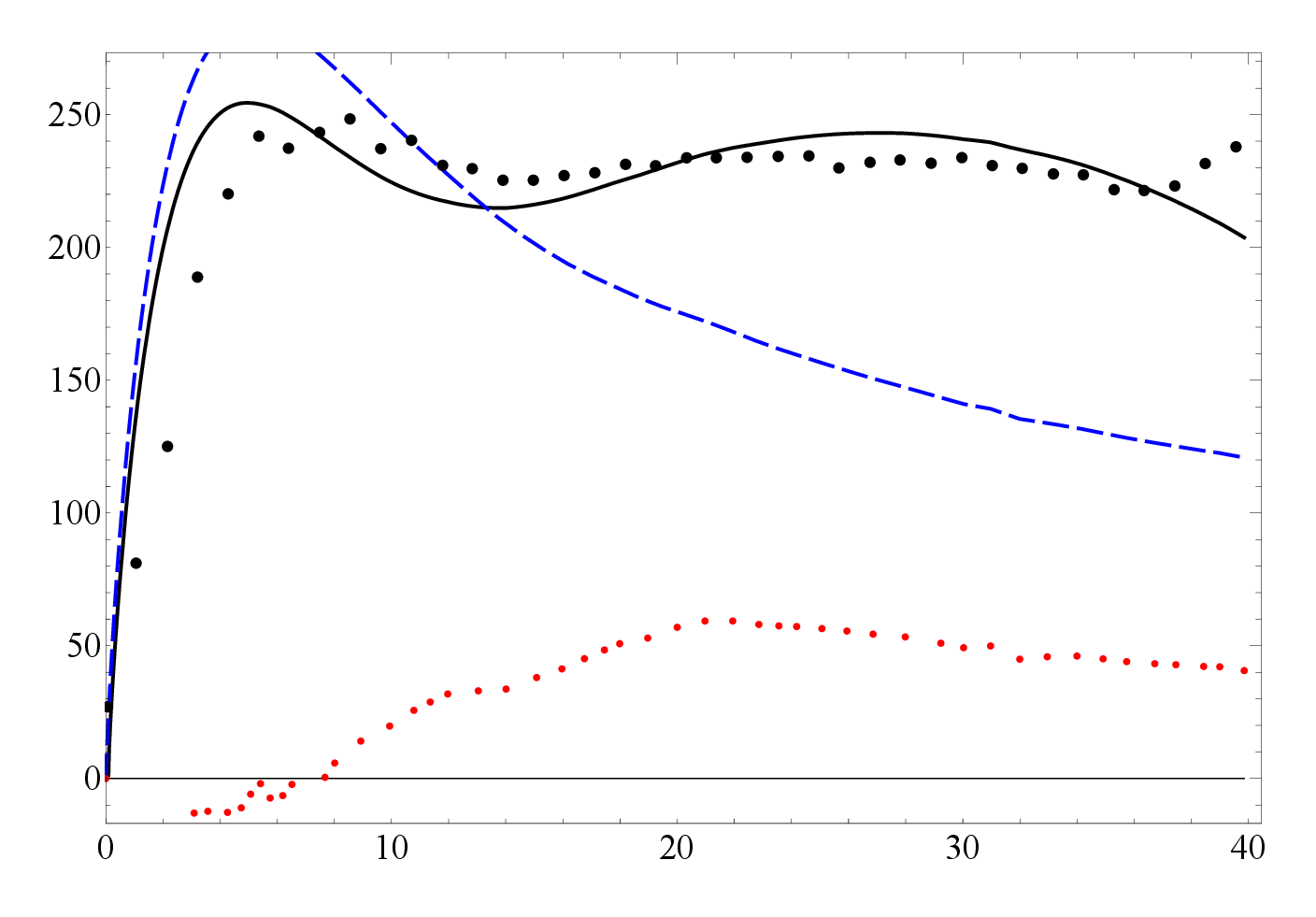}}
\subfloat[NGC 7793]{\includegraphics[width=\sc\columnwidth]{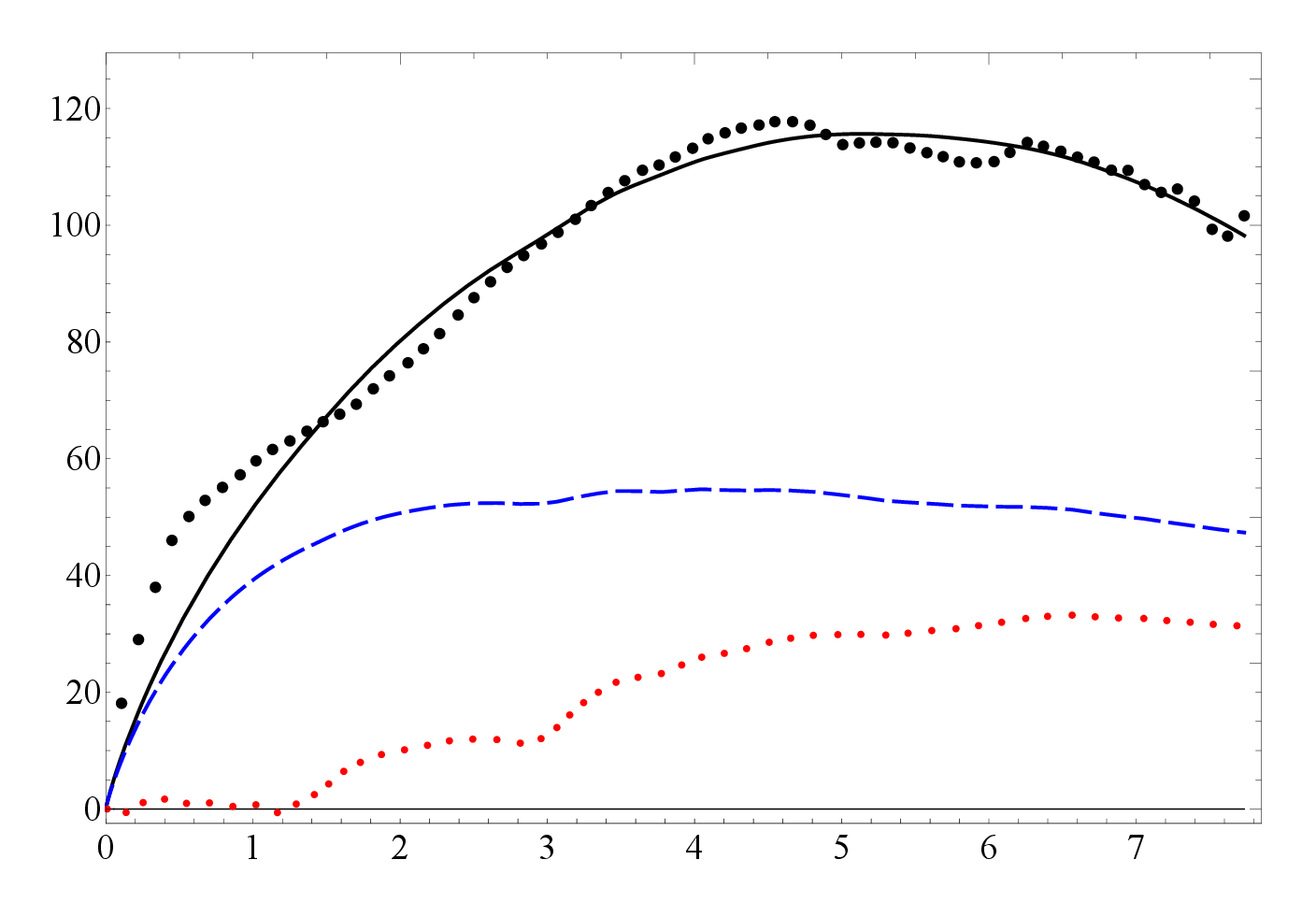}}
\caption{Least-squares fit (solid curves) to the rotation curves of sample galaxies,
for parameters listed in table \ref{t:fit}.
The horizontal axis is the distance $R$ in kpc, and the vertical axis is the rotation velocity in km s$^{-1}$.
Black dots are mean values from the THINGS data,
dotted curves are the contribution of gas, dashed curves are the total contribution of gas and stellar disc in Newtonian dynamics.
Cases of NGC 7331 and 7793 are separately discussed in \Figs \ref{f:diff} and \ref{f:diff2}, respectively.
}
\label{f:gal}
\end{figure}

\squeezetable
\begin{table}[h]
\centering
\resizebox{\textwidth}{!}{
\begin{tabular}{lrrrrrrrrrrr}
\hline
\multicolumn{1}{c}{Name}& \m D&\m{R_d}&\m{R_d}& \m{M_d}& \m{M_d}&\m{~~b_0/m}& \m{\pwr}&  \m{k_1}&    \m{k_2}&
\m{\tilde a_1}&\m{\tilde a_2}\\
\multicolumn{1}{c}{(1)} &\m{(2)}&\m{(3)}&\m{(4)}&\m{(5)}&\m{(6)}&      \m{(7)}&   \m{(8)}&\m{(9)}&\m{(10)}&
\m{(11)}&        \m{(12)}\\
\hline  
 & & & & & & & & & & \\ 
DDO 154                 &~~~ 4.3&~~   0.72&~~   0.54&~ 0.00263&~ 0.00186&~~~   $21.4^2$&~~~~~~        0&~~~~~~~   0&~~~~    0.233&~~~~~~~ 0&$~~ 4.12\times 10^{-36}$             \\
NGC  925                & 9.2&   3.30&   n/a&   1.02&   0.72&     326$^2$&    0.012&        0&  0.00174&
-4.34&$2.27\times 10^{-31}$             \\
NGC  2403               & 3.2&   1.81&  2.05&   0.47&   0.33&    111.5$^2$&        0&    0.171&        0&
-0.527&  $3.13\times 10^{-50}$             \\
NGC  2841               &14.1&   4.22&  3.55&  10.96&   7.59&      144$^2$&    0.050&        0&    0.089&
-2.39&  $5.14\times 10^{-50}$             \\
NGC  2903               & 8.9&   2.40&  2.81&   1.41&    1.0&     91.9$^2$&        0&        0&     0.18&
                     0&  $3.42\times 10^{-33}$             \\
NGC  2976               & 3.6&   0.91&   n/a&  0.178&  0.126&    111.4$^2$&        0&        0&    0.039&
                     0&  $4.44\times 10^{-31}$             \\
NGC  3031               & 3.6&   1.93&   n/a&   6.92&    4.9&    100.5$^2$&        0&        0&    0.056&
                     0&  $4.88\times 10^{-32}$             \\
NGC  3198               &13.8&   3.06&  3.88&   2.82&  1.995&     68.2$^2$&        0&        0&    0.017&
                     0&  $2.85\times 10^{-34}$             \\
NGC  3521               &10.7&   3.09&  2.86&   12.3&   8.71&      189$^2$&        0&        0&    0.017&
-7.03&  $1.68\times 10^{-47}$             \\
NGC  3621               &6.6 &   2.61&   n/a&   1.95&    1.38&    64.8$^2$&        0&    0.563&        0&
1.35&  $8.05\times 10^{-50}$             \\
NGC  4736               &4.7&   1.99 &   n/a&   1.86&    1.32&   230.4$^2$&        0.83&        0&    0.008&
-43.8&  $7.76\times 10^{-49}$             \\
NGC  5055               &10.1&   3.68&  5.00&   12.3&   8.71&     47.6$^2$&        0&     80.2&        0&
1.27&  $3.06\times 10^{-51}$             \\
NGC  6946               &5.9 &  2.97 &   n/a&   5.89&    4.17&     165$^2$&        0&      0.0543&    0&
                    0&  $1.72\times 10^{-46}$             \\
NGC  7331               &14.7&   2.41&  4.48&   16.6&  11.7&       311$^2$&        0&        0&   0.0015&
-20.1&  $1.25\times 10^{-49}$             \\
NGC  7793               &3.9 &  1.25 &   n/a&  0.275&  0.195&    91.6$^2$&        0&          0&    0.0436&
                    0&  $1.48\times 10^{-31}$             \\
 & & & & & & & & & & \\
\hline
\end{tabular}
} 
\caption{
The best fit of THINGS data, assuming formula \eqref{e:ov}.
Notes: 
(1) name of galaxy; 
(2) distance to galaxy, in Mpc; 
(3) characteristic radius of stellar disk taken from 
\cite{mc13}, in kpc; 
(4) characteristic radius of stellar disk taken from 
\cite{ll13}, in kpc; 
(5) mass of stellar disk according to the diet-Salpeter IMF, taken from 
\cite{mc13}, in $10^{10} M_\sun$;
(6) mass of stellar disk according to the Kroupa IMF, taken from 
\cite{mc13}, in $10^{10}M_\sun$;
(7) a free parameter, in km$^2$ s$^{-2}$;
(8) a free parameter, dimensionless;
(9) a free parameter, in kpc$^{-1}$;
(10) a free parameter, in kpc$^{-2}$;
(11) a free parameter (related to Rindler acceleration), in $10^{-11}$ m s$^{-2}$;
(12) a free parameter (related to de Sitter constant), in s$^{-2}$.
Values of free parameters were obtained assuming disk radius and mass from the third and sixth column, respectively.\\~
}
\lb{t:fit}
\end{table}

According to the popular exponential disk model \cite{fre70},
the rotation velocity of stars can be derived as 
\be\label{e:vstar}
  v_{\star}(R)
	=
	c_{(0)}
\left\{
\frac{M_d}{2 R_d} 
\left(
\frac{R}{R_d}
\right)^2
\left[
I_0\!\left(\frac{R}{2 R_d}\right) K_0\!\left(\frac{R}{2 R_d}\right)
- 
I_1\!\left(\frac{R}{2 R_d}\right) K_1\!\left(\frac{R}{2 R_d}\right)
\right]	
\right\}^{1/2}
	,
\ee
\ew
where $M_d$ and $R_d$ are the mass and length scale of the disk,
$I_{n}$ and $K_{n}$  being the $n$th-order modified Bessel functions of the first and second kind, respectively.

The Newtonian velocity due to the combined contributions of gas and stellar disk is given by
\be\label{e:vnwt}
 v_\text{N} (\ror) = \sqrt{
v_\text{gas}^2 (\ror)
+ v_{\star}^2 (\ror)
}
,
\ee
where the term $v_\text{gas}$ takes into account the contributions of both neutral hydrogen and helium. 

%
Furthermore, we perform the best fit to various galaxies data taken from 
the HI Nearby Galaxy Survey (THINGS)  \cite{wb08,bw08}.
We use formulae \eqref{e:ov}-\eqref{e:vnwt},
in which the fitting parameters are going to be
$b_0/m$,   $\pwr$,   $k_1$, $k_2$, $\tilde a_1$, $\tilde a_2$, whereas $R_d$ and  $M_d$
are
fixed according to an initial mass function model (IMF) chosen \cite{sa55,kr01},
as discussed in 
\cite{mc13,ll13}.
The gas contribution $v_\text{gas}$ can be directly deduced from the database,
similar plots can be found in 
\cite{bw08,ll13}. 
For simplicity, we use the mean values of the THINGS data.

\begin{figure}[t]
\centering
\subfloat[$R_d = 2.41$ kpc]{\includegraphics[width=\sct\columnwidth]{f_ngc7331.eps}}
\hspace{0mm}
\subfloat[$R_d = 3$ kpc]{\includegraphics[width=\sct\columnwidth]{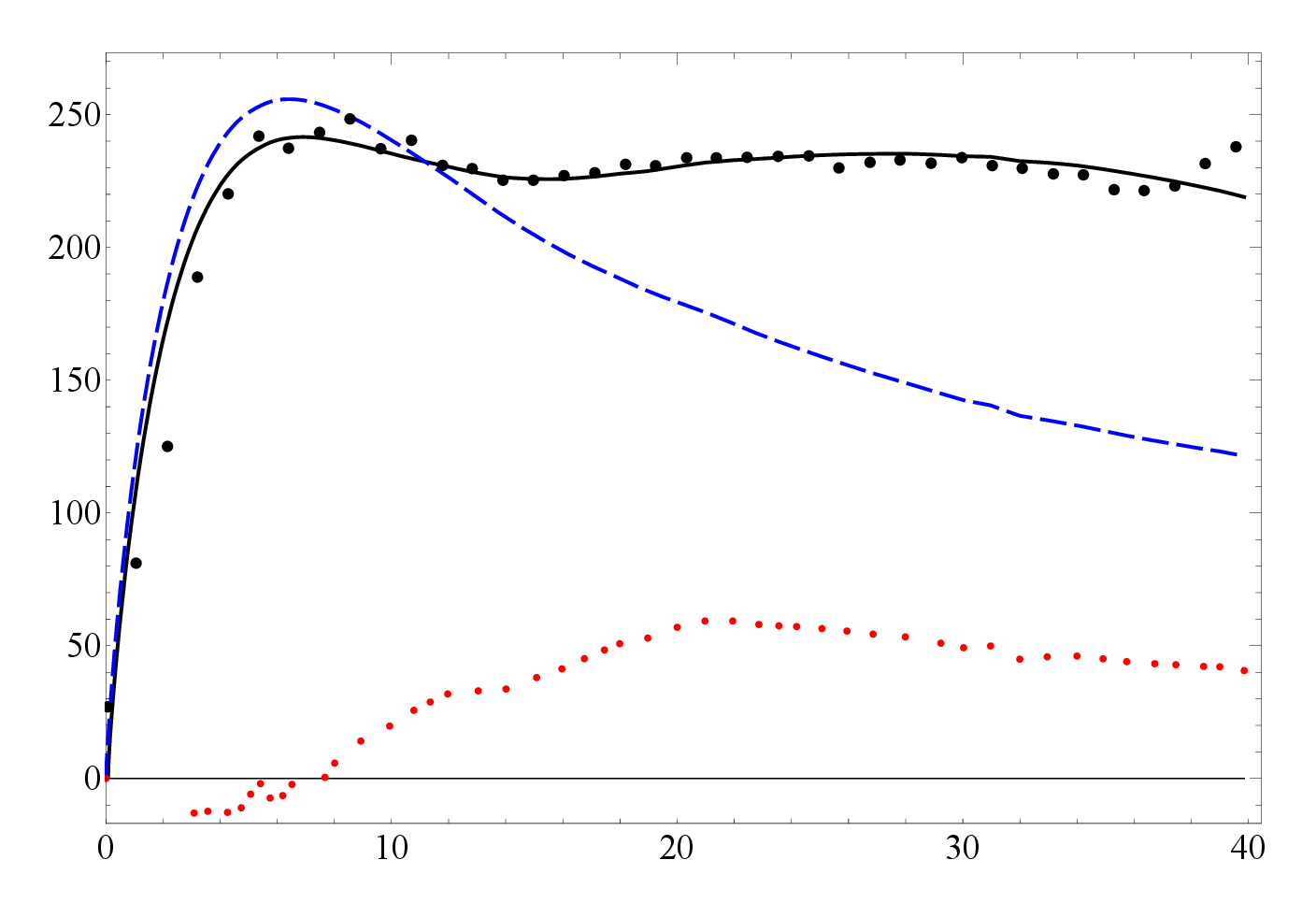}}
\caption{NGC 7331: a comparison of fits for different stellar disk parameters.
The first panel is the one taken from 
\Fig \ref{f:gal}, which is based on the fixed parameters from  third and sixth column of 
table \ref{t:fit};
fitting values of its free parameters are listed therein.
The second panel is the best fit based on the disk mass from the same sixth column, 
but the disk radius being assumed slightly larger; 
fitting values of the corresponding free parameters are:
$b_0/m = 238^2$ km$^{2}$ s$^{-2}$, 
$\pwr = 0$, $k_1 = 0$, 
$k_2 = 0.00147$ kpc$^{-2}$,
$\tilde a_1 = -1.02 \times 10^{-10}$ m s$^{-2}$
and
$\tilde a_2 = 1.15 \times 10^{-50}$ m s$^{-2}$.
Axes' labels and units are the same as in \Fig \ref{f:gal}.
 }
\label{f:diff}
\end{figure}

\begin{figure}[t]
\centering
\subfloat[$R_d = 1.25$ kpc]{\includegraphics[width=\sct\columnwidth]{f_ngc7793.eps}}
\hspace{0mm}
\subfloat[$R_d = 0.9$ kpc]{\includegraphics[width=\sct\columnwidth]{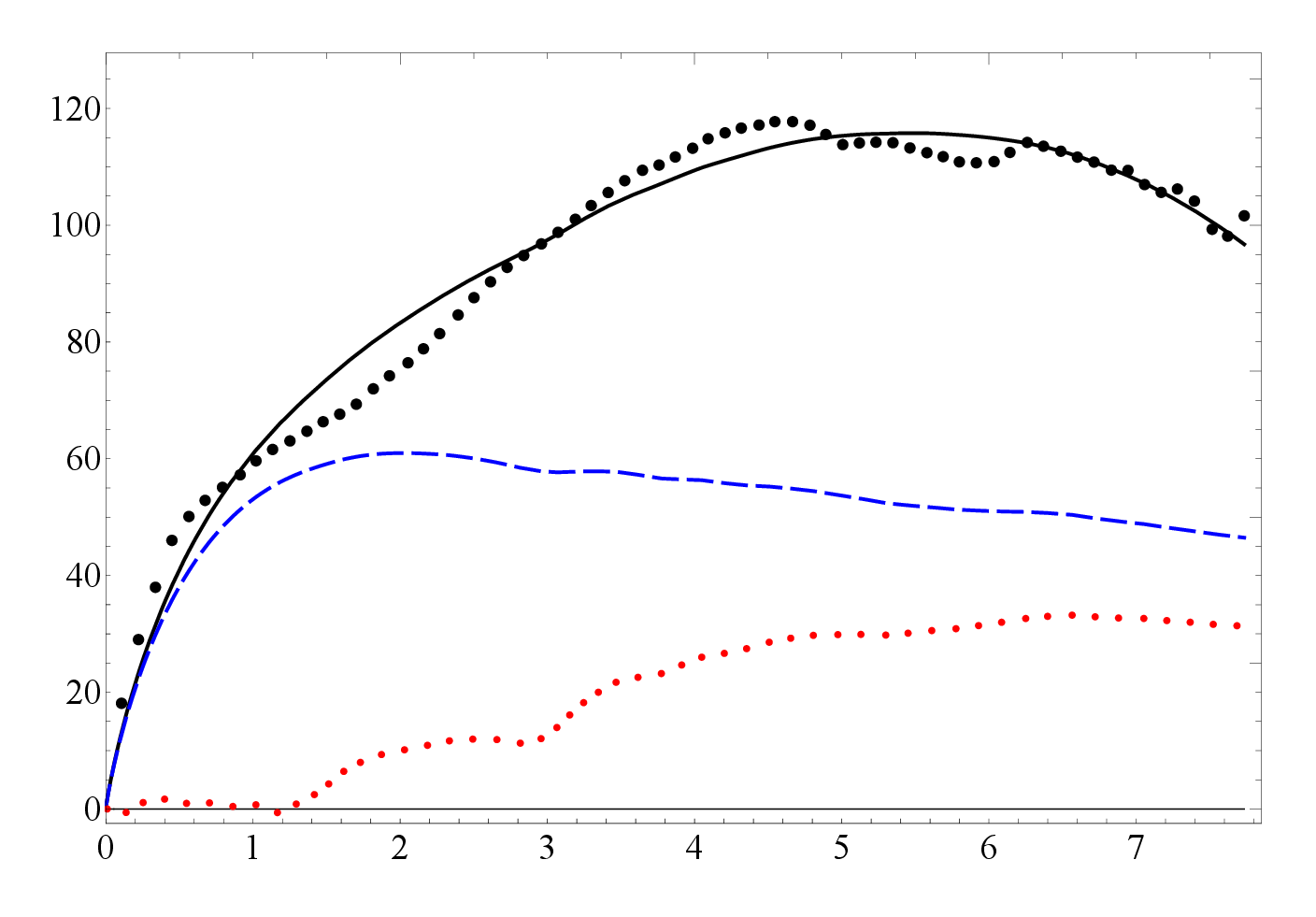}}
\caption{NGC 7793: a comparison of fits for different stellar disk parameters.
The first panel is the one taken from 
\Fig \ref{f:gal}, which is based on the fixed parameters from  third and sixth column of 
table \ref{t:fit};
fitting values of its free parameters are listed therein.
The second panel is the best fit based on the disk mass from the same sixth column, 
but the disk radius being assumed slightly smaller; 
fitting values of the corresponding free parameters are:
$b_0/m = 123^2$ km$^{2}$ s$^{-2}$, 
$\pwr = 0$, $k_1 = 0$, 
$k_2 = 0.0229$ kpc$^{-2}$,
$\tilde a_1 = 0$
and
$\tilde a_2 = 2.44 \times 10^{-31}$ m s$^{-2}$.
Axes' labels and units are the same as in \Fig \ref{f:gal}.
 }
\label{f:diff2}
\end{figure}

We make use of the Levenberg–Marquardt least-squares fit method.
The results are plotted in \Fig \ref{f:gal} 
(cases NGC 7331 and 7793 are being additionally discussed in \Figs \ref{f:diff} and \ref{f:diff2}),
and  
corresponding best-fitting parameters are
listed in table \ref{t:fit}.
When fitting, we fix the stellar disk radius and mass to the values from its third and sixth column, respectively.

\scn{Discussion}{s:dis}
Before discussing the fitting results themselves,
it should be stressed out 
that there is a large discrepancy in the interpretation of photometric data for stellar disks. 
Even within the common framework of Freeman's exponential disk model,
one gets substantially distinct values of stellar disk parameters, such as
mass and characteristic radii.
This uncertainty leads to the increased ambiguity of fitting curves, especially considering 
the multi-parameter character of the fitting function \eqref{e:ov}.
For these reasons, most of the fitting results are sensitive to the disk models used,
see \Figs \ref{f:diff} and \ref{f:diff2} for examples.
Yet, a few qualitative features,
common for sample galaxies considered
can be spotted right away.

To begin with, the power degree $\pwr$ is 
responsible for the part of the logarithmic term which diverges at $\ror \to 0$.
This parameter 
turns out to be zero for most of galaxies, except those
whose rotation velocity is not approaching small values at $\ror \to 0$,
such as NGC 925 or 2841.
It is not clear however if this occurs due to the lack of precise
observational data points at small orbital radii (less than 0.5 kpc).

One immediately notices that the logarithmic coupling $b_0/m$ 
acquires a large value for all 
fitting samples. 
On the other hand,
its square root is approximately equal to the average velocity
of the galaxy's FRC regime (if a plateau occurs on a curve),
or to the peak velocity (if a curve does not have a clearly visible plateau).
This suggests that the logarithmic term \eqref{e:grgal}
yields an important, if not predominant, contribution 
to galactic dynamics for a large range of distances, all the way up to the galactic outskirts.
This confirms our earlier result that the logarithmic term is largely responsible for 
the FRC regime's occurrence the outer regions of galaxies \cite{z20un1},
also that it contributes to the crossover from flat to non-flat regimes at galactic outskirts \cite{z21rc}.

Furthermore,
the linear term seems to affect the shapes of the rotation curves in middle-to-far regions, but in many cases the 
fitting result 
turns out better without assuming this term whatsoever -- which may or may not be a consequence of the stellar disk model's uncertainty mentioned above.
The quadratic (de Sitter) term contributes to asymptotic behaviour of rotation curves,
at the galaxy’s border.
Despite de Sitter constant's value looking rather small (in sub-astronomical units such as SI or CGS),
its presence substantially improves fitting at largest values of the orbital radius. 
 
It seems also that both linear and quadratic terms are unique to each galaxy, which confirms the conjecture of multiple expansion mechanisms, discussed in section 7 of 
\cite{z20un1};
the galaxy dependence of parameters of effective potentials (therefore, rotation velocities) was discussed in sections 2 and 6 thereof.
In essence, nearly all the parameters of the logarithmic superfluid vacuum theory are not the parameters \textit{per se},
but they are dynamical and thermodynamical values.
Therefore, they can vary, depending on the
environment and boundary conditions:
background superfluid not only induces gravitational potential, but also gets affected by it,
because this potential acts upon surrounding matter by creating density inhomogeneity.

\scn{Conclusion}{s:con}
In logarithmic superfluid theory, effective gravitational potential
is induced by the quantum wavefunction of the physical vacuum of a self-gravitating body or configuration;
while the vacuum itself is viewed as the superfluid described by the logarithmic wave equation in
the leading-order approximation with respect to Planck constant.
Therefore, predicted phenomena are determined not only by original parameters of the model
but also shape characteristics of the solution's wavefunction which also depend on boundary conditions.
This results in a rich mathematical structure of the theory.
For instance, gravity acquires a multiple-scale pattern,
to such an extent that
one can distinguish the
sub-Newtonian, Newtonian, logarithmic, linear (sometimes referred to as Rindler) and quadratic (de Sitter) terms in the effective potential.

Following the lines of our previous works \cite{z20un1,z21rc},
we have applied the best-fitting procedures to the rotation curve data of a number of galaxies from the HI Nearby
Galaxy Survey assuming their stellar disk and gas parameters being rigidly fixed to the mean values obtained 
using photometric methods.
Although the fitting results seem to be sensitive to stellar disk models and technical assumptions about the superfluid vacuum's wavefunction, they agree with observational data very well, 
considering the large range of phenomena involved.

It should be emphasized that 
successful fitting also takes place for those galaxies whose rotation curves do not have a 
clearly visible FRC plateau, such as NGC 2976, 4736 and 7793, 
or have unusual asymptotic behaviour at small values of the orbital radius, such as NGC 925, 2841 and 4736.
This is largely possible due to the multi-scale structure of gravity in logarithmic superfluid vacuum theory.

\begin{acknowledgments}
Proofreading of the manuscript by P. Stannard is greatly appreciated.
This research was funded by Department of Higher Education and Training
of South Africa
and in part by National Research Foundation of South Africa (Grants Nos. 95965, 131604 and 132202).
\end{acknowledgments}

\end{document}